\documentclass[conference,10pt,a4paper]{IEEEtran}
\IEEEoverridecommandlockouts

\makeatletter
\def\markboth#1#2{\def\leftmark{\@IEEEcompsoconly{\sffamily}\MakeUppercase{\protect#1}}%
\def\rightmark{\@IEEEcompsoconly{\sffamily}\MakeUppercase{\protect#2}}}
\makeatother

 \PassOptionsToPackage{bookmarks={false}}{hyperref}
\usepackage[utf8x]{inputenc}
\usepackage[english]{babel}
\usepackage{ucs}
\usepackage{amsmath}
\usepackage{amsfonts}
\usepackage{amssymb}
\usepackage{amsthm}
\usepackage{array}
\usepackage{verbatim}
\usepackage{listings}
\usepackage{psfrag}
\usepackage{stfloats}

\usepackage{algorithm}
\usepackage{algorithmic}
\usepackage{url}
  \usepackage{enumerate}
  \usepackage{cite}
\usepackage[usenames,dvipsnames,svgnames,table]{xcolor}
\usepackage[margin=0.75in]{geometry}

\ifCLASSINFOpdf
  \usepackage[pdftex]{graphicx}
  \usepackage{subfigure}
  \usepackage{epstopdf}
  \graphicspath{{./img/}}
   \DeclareGraphicsExtensions{.eps,.ps,.png,.pdf}
\else
  \usepackage[dvipdf]{graphicx}
  \usepackage{subfigure}
  \graphicspath{{./img/}}
   \DeclareGraphicsExtensions{.eps,.ps}
\fi

\newcommand{\F}{\mathbf{F}}

\newcommand{\A}{\mathbf{A}}

\newcommand{\D}{\mathbf{D}}

\newcommand{\I}{\mathbf{I}}

\newcommand{\Pb}{\mathbf{P}}
\newcommand{\Phib}{\mathbf{\boldsymbol{\Phi}}}

\newcommand{\h}{\mathbf{h}}
\newcommand{\x}{\mathbf{x}}

\newcommand{\vv}{\mathbf{v}}

\newcommand{\y}{\mathbf{y}}
\newcommand{\bb}{\mathbf{b}}

\newcommand{\rr}{\mathbf{r}}
\newcommand{\z}{\mathbf{z}}
\newcommand{\ab}{\mathbf{a}}
\newcommand{\pp}{\mathbf{p}}

\newcommand{\tr}{\textnormal{tr}}

\newcommand{\Ex}[2]{{\textnormal{E}_{#1}\left[#2\right]}}

\usepackage{pifont}
\newcommand\Tau{\mathcal{T}}

\theoremstyle{plain}

   \definecolor{blueH3}{rgb}{0,.5,1}
   \definecolor{blueH2}{rgb}{0,0.25,0.75}
   \definecolor{blueH1}{rgb}{0,0,0.5}   
   \definecolor{grayOldText}{rgb}{.5,.5,.5}
   \definecolor{VCobalt}{HTML}{005682}
   \definecolor{TZTeal}{HTML}{008080}
   \definecolor{KYJade}{HTML}{008151}
   \definecolor{ARust}{HTML}{a10000}
   \definecolor{FFucsia}{HTML}{7000c3}
   
\newcommand{\CASE}[1]{\STATE \textbf{case} #1\textbf{:} \begin{ALC@g}}
\newcommand{\ENDCASE}{\end{ALC@g}}

\newcommand{\DEFAULT}{\STATE \textbf{default:} \begin{ALC@g}}
\newcommand{\ENDDEFAULT}{\end{ALC@g}}
\newcommand{\DEFAULTLINE}[1]{\STATE \textbf{default:} }

\begin{document}
\sloppy

\title{Sparse mmWave OFDM Channel Estimation Using Compressed Sensing in OFDM Systems.}

  \author{
  \IEEEauthorblockN{Felipe Gomez-Cuba$^1$ $^2$; Andrea J. Goldsmith$^1$}
  \IEEEauthorblockA{
    $^1$Electrical Engineering, Stanford University, Stanford, CA,\\
    $^2$Dipartimento D'engegneria della Informazione, University of Padova, Italy\\
    { Email: \texttt{\{gmzcuba,andreag\}@stanford.edu } } }
}

\maketitle

\begin{abstract}

This paper proposes and analyzes a mmWave sparse channel estimation technique for OFDM systems that uses the Orthogonal Matching Pursuit (OMP) algorithm. This greedy algorithm retrieves one additional multipath component (MPC) per iteration until a stop condition is met. We obtain an analytical approximation for the OMP estimation error variance that grows with the number of retrieved MPCs (iterations). The OMP channel estimator error variance outperforms a classic maximum-likelihood (ML) non-sparse channel estimator by a factor of approximately $2\hat{L}/M$ where $\hat{L}$ is the number of retrieved MPCs (iterations) and $M$ the number of taps of the Discrete Equivalent Channel. When the MPC amplitude distribution is heavy-tailed, the channel power is concentrated in a subset of \textit{dominant} MPCs. In this case OMP performs fewer iterations as it retrieves only these dominant large MPCs. Hence for this MPC amplitude distribution the estimation error advantage of OMP over ML is improved.
In particular, for channels with MPCs that have lognormally-distributed amplitudes, the OMP estimator recovers approximately 5-15 dominant MPCs in typical mmWave channels, with 15-45 weak MPCs that remain undetected.
\end{abstract}


\section{Introduction}
As demand for higher data rates continues to grow, the wireless industry is looking to support it through increased bandwidth transmissions. 
Tens of GHz are available in the mmWave spectrum ($30-300$GHz) \cite{Mathew2016}. Extensive measurements indicate that mmWave channels experience \textit{sparse scattering} \cite{Mathew2016,Specification2017} due to severe attenuation or blocking of most reflections. The few multipath reflections that reach the receiver have arrival delays spread over a relatively long time interval. Orthogonal Frequency Division Multiplexing (OFDM) systems separate the frequency-selective channel into a series of frequency-flat subcarriers with scalar gains 
\cite{Morelli2001}. Since only a few multipath components (MPC) with resolvable delays exist, this channel is ``frequency-selective'' but does not exhibit ``rich scattering''. Since there are significantly fewer MPCs than subcarriers, the OFDM channel can be written in a reduced number of dimensions as a linear combination of unitary column vectors forming a certain basis. 

Multipath channel models at frequencies below 6GHz typically assume a complex Gaussian distribution for the amplitudes of each MPC. This is justified for OFDM systems when the signal bandwidth of the subcarrier is narrowband and there are sufficiently many MPCs to invoke the Central Limit Theorem (CLT) \cite{goldsmith2005book}. Channel estimation in OFDM systems without sparsity has been extensively studied, for example in \cite{Morelli2001}. 

Compressive Sensing (CS) is a general framework for the estimation of sparse vectors under a linear observation matrix \cite{Duarte2011}. CS methods have been applied to sparse channel estimation, including OFDM systems in sub-6GHz frequency bands \cite{Taubock2008,Berger2010a,Qi2011a,Meng2012} as well as mmWave systems in the time domain \cite{Mo2017,7953407}, OFDM \cite{rodriguez2017frequency} and joint time-frequency domain \cite{Venugopal2017}. In sub-6GHz bands the MPC amplitudes are assumed to follow a hybrid Bernoulli-Gaussian random vector distribution \cite{Taubock2008,Berger2010a,Qi2011a,Meng2012}. This model has also been used for mmWave channels in \cite{Mo2017,rodriguez2017frequency}. On the other hand the CS mmWave channel estimation models in \cite{7953407,Venugopal2017} did not discuss the MPC amplitude distribution or analyze its effect in performance, giving only
performance results determined by simulation.

Classic CS results often require that the estimated vector is of a specific form as a sufficient condition. For example in \cite[Theorems 6,7]{Duarte2011} error bounds are guaranteed if the estimated vector is strictly sparse with a specific number of non-zeros. Nevertheless, the relation between CS estimation performance and random sparse vector distributions is not yet fully understood. A recent advance in this area in \cite{Gribonval2012} established that the ``compressibility'' of a random i.i.d. vector distribution depends on the second and fourth moments. However the assumptions in \cite{Gribonval2012} are not directly applicable to a practical mmWave OFDM channel estimation scenario. This is because the results in  \cite{Gribonval2012} are asymptotic in the limit as the number of i.i.d. vector dimensions goes to infinity, and under a CS problem that is slightly different from OFDM channel estimation.

The NYUWireless center \cite{Mathew2016} and 3GPP New Radio specification \cite{Specification2017} have published measurement-based models for the generation of random synthetic mmWave sparse multipath channels. Both of these models use a heavy-tailed lognormal distribution to model MPC amplitudes. Thus, these models generate random channel vectors with different sparsity characteristics than the Bernoulli-Gaussian distribution assumed in prior CS OFDM sparse channel estimation works \cite{Taubock2008,Berger2010a,Qi2011a,Meng2012,Mo2017,7953407,rodriguez2017frequency,Venugopal2017}. Since distribution heavy-tailed-ness (fourth moment) is connected to CS performance  \cite{Gribonval2012}, if we synthesize random mmWave channels following the generator models established in \cite{Mathew2016,Specification2017} the performance of CS OFDM channel estimators will be different from the simulation results of \cite{7953407,rodriguez2017frequency,Mo2017,Venugopal2017}. As the models in \cite{Mathew2016,Specification2017} are based on mmWave measurements in outdoor dense urban microcell scenarios, this means the performance results of \cite{7953407,rodriguez2017frequency,Mo2017,Venugopal2017} may not be realistic in these scenarios.

In this paper we analyze the performance of a practical OMP CS algorithm and study the influence that the random MPC amplitudes has on this performance. We consider a finite observation matrix that is the DFT of a sparse frequency-selective channel with arbitrary multipath delays. In CS the term \textit{superresolution} describes sensing matrices that enable vector reconstruction surpassing the Shannon-Nyquist sampling limits \cite{Duarte2011}. A low Mutual Incoherence (MI) and the Restricted Isometry Property (RIP) are defined in \cite{Duarte2011} as the sufficient conditions for classic CS error analyses. Unfortunately, unlike Gaussian matrixes, a DFT sensing matrix with superresolution does not have these properties \cite{Berger2010a}. Our main contribution is an approximation of the error variance of OMP that does not require neither low MI or the RIP for the proof. We show the error of our OMP estimation method grows linearly with the number of iterations and the approximation is tight in the high-SNR regime.

Based on this error analysis we characterize the ``compressibility'' of the channel through a decaying function $\overline\rho(d)$ that measures the residual channel power that is not accounted for when OMP retrieves $d$ MPCs. The faster $\overline\rho(d)$ decays, the fewer iterations OMP performs, and the lower the error. In order to gauge the decay-speed of $\overline\rho(d)$ we introduce the Fairness Index \cite{Jain1984} as a measure of the inequality of arbitrary vectors. The FI of a random channel is a random variable connected to the heavy-tailed-ness of the distribution. The mmWave random channel generators in \cite{Mathew2016,Specification2017} use a heavy-tailed lognormal, have lower FI, and are more compressible than Bernoulli-Gaussian vectors. Therefore OMP OFDM channel estimation for mmWave retrieves fewer MPCs and has a lower error variance than in prior works that assumed Bernoulli-Gaussian MPCs.

For the sake of the ease of exposition, we present our results for a single antenna system model where only delay-domain sparse channel estimation is considered. We believe our analysis can be extended to the joint angular and delay sparsity CS estimation schemes for MIMO mmWave systems presented in \cite{7953407,rodriguez2017frequency,Mo2017,Venugopal2017}. We
leave the extension of our results to multiple-antenna models for future work.

The rest of the paper is structured as follows: 
Sec. \ref{sec:model} describes the system model. Sec. \ref{sec:LSML} defines two Maximum Likelihood (ML) estimator benchmarks. Sec. \ref{sec:CS} describes the OMP channel estimator. Sec. \ref{sec:OMPstop} contains the random vector compressibility analysis. Sec. \ref{sec:num} provides numerical results.
Sec. \ref{sec:conclusions} concludes the paper.

\section{System Model}
\label{sec:model}
\subsection{Multipath Time-Domain Channel Model}

We consider a time-invariant discrete-time equivalent channel (DEC) with Finite Impulse Response (FIR) length $M$ as in \cite{rodriguez2017frequency,7953407,Mo2017,Venugopal2017}. The channel is the sum of $L$ planar waves with fixed amplitudes $\{\alpha_{\ell}\}_{\ell=1}^{L}$, phases $\{\phi_{\ell}\}_{\ell=1}^{L}$ and delays $\{\tau_{\ell}\}_{\ell=1}^{L}$. The discrete-time conversion is modeled by a transmit pulse $p(t)$ and the sampling period $T$.
\begin{equation}
 \label{eq:chanM}
 h_M[n]=\sum_{\ell=1}^{L}\alpha_{\ell}e^{j\phi_{\ell}}p(nT-\tau_{\ell}),\quad n\in[0,M-1]
\end{equation}

The set of delays $\{\tau_{\ell}\}_{\ell=1}^{L}$ is ordered ($\tau_\ell>\tau_{\ell-1}$) and aligned to zero ($\tau_1=0$). The maximum delay spread is $D_s=\max \tau_{\ell}=\tau_L$. Typical choices of $p(t)$ have a peak at $t=0$ and weak infinite tails, so choosing $M= \lceil D_sT\rceil$ guarantees that all the MPCs are contained in the FIR DEC. 

We assume $L\leq M$, generate the sets $\{\alpha_{\ell}\}_{\ell=1}^{L}$, $\{\phi_{\ell}\}_{\ell=1}^{L}$ and $\{\tau_{\ell}\}_{\ell=1}^{L}$ randomly, and apply \eqref{eq:chanM} to obtain the channel. For this model $h_M[n]$ has a probability density function that is too cumbersome to write explicitly. For typical pulses $p(t)$, the more heavy tailed the distribution of $\{\alpha_{\ell}\}_{\ell=1}^{L}$, the more unevenly the total energy transfer of the channel is distributed among the taps $h_M[n]$.

We assume the carrier frequency is $f_c>10/T$, the delay separations satisfy $\tau_\ell-\tau_{\ell-1}\gg 1/f_c=\lambda/c$, and the phases $\{\phi_{\ell}\}_{\ell=1}^{L}$ are independent and uniformly distributed in the range $[0,2\pi)$. This assumption is a key difference between mmWave \cite{Mathew2016,Specification2017} and sub-6GHz ultra-wideband channels.

The MPC delays $\{\tau_{\ell}\}_{\ell=1}^{L}$ in our model follow a Poisson Arrival Process (PAP). 
Usually delays are modeled by ``clustering'' in groups forming a Non-Uniform PAP (NUPAP). Different references use different criteria to define clusters but this does not affect our analysis \cite{Mathew2016,Specification2017}.

The MPC amplitudes $\{\alpha_{\ell}\}_{\ell=1}^{L}$ follow a normalized lognormal distribution with a delay-dependent mean, where the unnormalized amplitudes satisfy $\log \overline{\alpha}_{\ell} = -\tau_\ell/\Gamma +\zeta_\ell$. Here $\Gamma$ is the mean received power decay with the delay and $\zeta_\ell\sim  \mathcal{N}(0,\sigma_\alpha^2)$ is a shadowing distribution that randomizes the amplitude. 
The normalization $\alpha_\ell=\sqrt{\frac{P_{recv}}{\sum_{\ell=1}^{L} \overline{\alpha}_\ell^2}}\overline{\alpha}_\ell$ where $P_{recv}$ is the total received power is applied for consistency with existing macroscopic shadowing and pathloss.  When the delay NUPAP features clustering, this makes both the delays \textit{and the amplitudes} of the MPCs dependent on each other. This is modeled in \cite{Mathew2016,Specification2017} with two lognormals: one for clusters and one for subpaths.

For convenient notation we rewrite \eqref{eq:chanM} as a vector. First, we define the size-$M$ time-domain channel vector
$$\h_M\triangleq\left(h_M[0],h_M[1],h_M[2],\dots,h_M[M-1]\right)^T.$$
Second, we define the size-$M$ pulse-delay vector
$$\pp(\tau)\triangleq\left(p(-\tau),p(T-\tau),\dots,p((M-1)T-\tau)\right)^T$$
where typical digital communication systems are implemented with a pulse $p(t)$ such that $|\pp(\tau)|^2\simeq1$ and if $\tau_\ell-\tau_{\ell-1}>T/2$ then $\pp(\tau_\ell)^H\pp(\tau_{\ell-1})\ll1$. Third, we define the size-$M\times L$ pulse-delay matrix
$$\Pb_{\{\tau_\ell\}_{\ell=1}^{L}}\triangleq\left(\pp(\tau_1),\pp(\tau_2),\dots,\pp(\tau_L)\right)$$
where, for typical pulses $p(t)$, we have $||\Pb_{\{\tau_\ell\}_{\ell=1}^{L}}||^2\simeq L$ and if $\tau_\ell\neq\tau_{\ell'}\forall \ell\neq\ell'$ then $\Pb_{\{\tau_\ell\}_{\ell=1}^{L}}$ is full rank. And finally, we define the size-$L$ MPC complex gain vector
$$\ab\triangleq\left(\alpha_1e^{j\theta_1},\alpha_2e^{j\theta_2},\alpha_3e^{j\theta_3},\dots,\alpha_Le^{j\theta_L}\right)^T.$$
Using these definitions, the sum expressed in \eqref{eq:chanM} can be equivalently written as the following matrix expression
\begin{equation}
 \label{eq:chanPb}
 \h_M=\sum_{\ell=1}^{L}\pp(\tau_\ell)\alpha_\ell e^{j\phi_\ell}=\Pb_{\{\tau_\ell\}_{\ell=1}^{L}}\ab.
\end{equation}


\subsection{OFDM Channel Model and Pilot Scheme}


We assume an OFDM digital system where the transmitter concatenates the Inverse DFT (IDFT) of $K\geq M$ data coefficients and $M$ Cyclic Prefix (CP) coefficients and sends the resulting frame through the FIR DEC (1). After the receiver discards the CP and computes the DFT
the frequency-domain channel is
\begin{equation}
 \label{eq:chanOFDM}
 \y=\D(\x)\h_K+\z,
\end{equation}
where all vectors have $K$ coefficients: $\z$ is Additive White Gaussian Noise (AWGN) with variance $\sigma^2\I_K$, $\x$ are the $K$ channel inputs, $\D(\x)$ is the diagonal-matrix containing the vector $\x$ in the main diagonal, and $\h_K$ is the frequency-domain channel obtained by computing the size-$K$ DFT of \eqref{eq:chanPb}.
For notational compactness, we can ignore the last $K-M$ columns of the DFT matrix, writing
\begin{equation}
 \label{eq:chanK}
 \h_K=\F_{K,M} \h_M
\end{equation}
where the rectangular matrix $\F_{K,M}$ contains the first $M$ columns of the DFT and satisfies $\F_{K,M}^H\F_{K,M}=\I_M$ but $\F_{K,M}\F_{K,M}^H$ is not an identity matrix.

Since \eqref{eq:chanM} is assumed time-invariant, so is $\h_K$. In practical scenarios this assumption is adopted when the phases $\{\phi_{\ell}\}_{\ell=1}^{L}$ do not change significantly over the duration of a block of several OFDM frames. In the first OFDM frame of each block, some subcarriers are selected to transmit a pilot pattern that is known to the receiver, which estimates the channel using the pilots before data reception. We denote by $N$ the number of pilot subcarriers where $K\geq N\geq M$, and we assume the pilot subcarriers are regularly spaced among the data subcarriers following a ``comb'' pilot structure. That is, $K/N$ is an integer so for example if $K/N=3$ then the subcarrier indexes $0,3,6\dots$ correspond to pilots symbols.

We denote the size-$N$ vector of transmitted pilots by $\x_N$ (a subset of the frequency-domain OFDM symbol $\x$ where pilots are transmitted). We define the vector of received coefficients in subcarriers corresponding to pilots during the first OFDM frame as follows:
$$\y_N=\D(\x_N)\h_{N/K}+\z_N.$$
Denoting by $\F_{N/K,M}$ the submatrix that contains the \textit{first} $M$ columns and the \textit{alternated} $N$\textit{-out-of-}$K$ rows of the DFT matrix, the channel for the pilots is
\begin{equation}
 \label{eq:chanN}
 \h_{N/K}=\F_{N/K,M}\h_M
\end{equation}
and with $N\geq M$ the full channel $\h_K$ can be recovered substituting \eqref{eq:chanK} in \eqref{eq:chanN} to write $\h_K=\F_{K,M}\h_M=\F_{K,M}\F_{N/K,M}^H\h_{N/K}$. 

Substituting \eqref{eq:chanPb} into \eqref{eq:chanK} we can write two alternative linear representations of the OFDM channel $\h_K$ as follows.
\begin{equation}
 \label{eq:chanMPClinear}
 \underset{\textnormal{frequency}}{\underbrace{\h_K}}=\underset{\textnormal{discrete-time}}{\underbrace{\F_{K,M}\h_M}}=\underset{\textnormal{MPCs}}{\underbrace{\F_{K,M}\Pb_{\{\tau_\ell\}_{\ell=1}^{L}}\ab}},
\end{equation}
Using these two linear representations, non-sparse and sparse channel estimators can be written, respectively, as we will describe in more detail in the next section.

\section{LS/ML Estimation Benchmarks}
\label{sec:LSML}


\subsection{Conventional Discrete-time-domain LS/ML estimation}
\label{sec:LSMLconventional}
A non-sparse ML estimator of $\h_K$ is given in \cite{Morelli2001}, using $\h_K=\F_{K,M}\h_M$ but not requiring that $\h_M$ is sparse. This non-Bayesian estimator is the best we can do in non-sparse channels when the distribution of	 $\h_M$ is unknown or untractable, and we adopt it as the ``non-sparse benchmark''. 

%
%

For $M\leq N \leq K$ it can be shown that the ML estimator of $\h_K$ subject to a linear constraint $\h_K=\F_{K,M}\h_M$ is a Least Squares (LS) estimator of $\h_M$ multiplied by $\F_{K,M}$ to reconstruct the frequency-domain vector $\h_K$ \cite{Morelli2001}:

\begin{equation}
\label{eq:MLM}
\hat{\h}_M^{\textnormal{ML-}M}=(\D(\x_N)\F_{N/K,M})^\dag\y_N,
\end{equation}
\begin{equation}
\label{eq:MLMK}
\hat{\h}_K^{\textnormal{ML-}M}=\F_{K,M}\hat{\h}_M^{\textnormal{ML-}M},
\end{equation}
where $\A^\dag$ is the Moore-Penrose left-side pseudoinverse, $(\A^H\A)^{-1}\A^H$, and the error can be expressed as
$$\tilde{\h}_K^{\textnormal{ML-}M}\triangleq\hat{\h}_K^{\textnormal{ML-}M}-\h_K=\F_{K,M}(\D(\x_N)\F_{N/K,M})^\dag\z_N.$$
Since $\z_N$ is AWGN the error is Gaussian and has variance
\begin{equation}
\label{eq:errMLMK}
\begin{split}
\nu_{\textnormal{ML-}M}^2&=\frac{\Ex{\z}{|\F_{K,M}(\D(\x_N)\F_{N/K,M})^\dag\z_N|^2}}{K}
 \geq\frac{M}{N}\sigma^2,
\end{split}
\end{equation}
where the equality is achieved by pilot sequences with unit-amplitude coefficients $|x[k]|=1\forall k$. In our numerical simulations, we verified this error variance result (Fig. \ref{fig:mse128}).

\subsection{Sparse MPC-domain LS/ML estimation}
\label{sec:LSMLMPC}

We define our second benchmark assuming the channel is sparse and $\{\tau_\ell\}_{\ell=1}^{L}$ is perfectly known to the receiver (as if ``revealed by a genie''). 
With minor changes to Sec. \ref{sec:LSMLconventional} we can derive a LS estimator of the vector $\ab$ and use $\F_{K,M}\Pb_{\{\tau_\ell\}_{\ell=1}^{L}}$ to recover the frequency-domain channel as
\begin{equation}
\label{eq:MLa}
\hat{\ab}^{\textnormal{ML-}{\{\tau_\ell\}_{\ell=1}^{L}}}= (\D(\x_N)\F_{N/K,M}\Pb_{\{\tau_\ell\}_{\ell=1}^{L}})^{\dag}\y_N
\end{equation}
\begin{equation}
\label{eq:MLaK}
\hat{\h}_K^{\textnormal{ML-}{\{\tau_\ell\}_{\ell=1}^{L}}}=\F_{K,M}\Pb_{\{\tau_\ell\}_{\ell=1}^{L}}\hat{\ab}^{\textnormal{ML-}{\{\tau_\ell\}_{\ell=1}^{L}}}
\end{equation}

For AWGN $\z_N$ the error is Gaussian with variance
\begin{equation}
\begin{split}
\label{eq:errMLaK}
\nu_{\textnormal{ML-}{\{\tau_\ell\}_{\ell=1}^{L}}}^2
&\geq \frac{L}{N}\sigma^2,\\
\end{split}
\end{equation}
where the gain versus the non-sparse benchmark is $M/L$.

\section{Hybrid CS/LS estimator of Sparse MPCs}
\label{sec:CS}

\subsection{Overal Description}
The LS estimator of $\ab$ when $\Pb_{\{\tau_\ell\}_{\ell=1}^{L}}$ is perfectly known is not practical in the sense that the pulse $p(t)$ is known but the delays $\{\tau_\ell\}_{\ell=1}^{L}$ are not known by the receiver. Therefore $\Pb_{\{\tau_\ell\}_{\ell=1}^{L}}$  cannot be perfectly known. To design a practical channel estimator, we assume the channel is sparse in the sense that a linear representation $\h_K=\F_{K,M}\Pb_{\{\tau_\ell\}_{\ell=1}^{L}}\ab$ exists but its matrix is unknown, and we define a hybrid two-step estimator of the channel as follows:
\begin{enumerate}
 \item We use the OMP algorithm to estimate the delays of the channel, denoted as $\hat{\Tau}\simeq \{\tau_\ell\}_{\ell=1}^{L}$. This approximate delay set defines a base for a vector subspace $\mathcal{S}(\F_{K,M}\Pb_{\hat{\Tau}})$. We denote the projection of $\h_K$ in this space by $\h_S=\F_{K,M}\Pb_{\hat{\Tau}}(\F_{K,M}\Pb_{\hat{\Tau}})^\dag\h_K$. Since $\h_S\in\mathcal{S}(\F_{K,M}\Pb_{\hat{\Tau}})$, there exists some vector $\bb\in\mathbb{C}^{|\hat{\Tau}|}$ such that $\h_S=\Pb_{\hat{\Tau}}\bb$.
 \item We use MPC-domain estimation \eqref{eq:MLa} to estimate $\bb$:
 \begin{equation}
 \label{eq:bML}
 \hat{\bb}^{ML-\hat{\Tau}}=(\D(\x_N)\F_{N/K,M}\Pb_{\hat{\Tau}})^{\dag}\y_N
 \end{equation}
 and obtain an estimator of $\h_S$ using \eqref{eq:MLa}, i.e. 
 $$\hat{\h}_S^{ML-\hat{\Tau}}=\F_{N/K,M}\Pb_{\hat{\Tau}}\hat{\bb}^{ML-\hat{\Tau}}.$$
 
\end{enumerate}
Since $\h_S$ is a vector that is ``close'' to $\h_K$ per step 1, we declare $\hat{\h}_S^{ML-\hat{\Tau}}$ in step 2 to be an estimator of $\h_K$, where two orthogonal errors are introduced. In the first step $\h_{E}=\h_K-\h_S=\F_{K,M}(\I_M-\Pb_{\hat{\Tau}}\Pb_{\hat{\Tau}}^\dag)\F_{K,M}^H\h_K$ by definition is orthogonal to $\mathcal{S}(\F_{K,M}\Pb_{\hat{\Tau}})$. In the second step $\tilde{\h}_S=\h_S-\hat{\h}_S=\F_{K,M}\Pb_{\hat{\Tau}}(\bb-\hat{\bb})$ by definition is contained in $\mathcal{S}(\F_{K,M}\Pb_{\hat{\Tau}})$.
Thus the total error $\tilde{\h}_K=\h_K-\hat{\h}_S=\h_E+\tilde{\h}_S$ satisfies $|\tilde{\h}_K|^2=|\h_E|^2+|\tilde{\h}_S|^2$.

We assume that the pilot sequence $\x_N$ contains unit-amplitude symbols as was optimal in the LS amplitude estimators in the previous section. Without loss of generality we set $\D(\x_N)=\I_N$ where results remain the same except for a phase rotation using other pilot sequences.

\subsection{OMP EStimator of $\{\tau_\ell\}_{\ell=1}^{L}$}

The \textit{delay dictionary set} with size $N_T\geq M$ is defined as $\Tau_{N_T}\triangleq\{n\frac{D_s}{N_T}:n\in[0,N_T-1]\}$. The \textit{delay dictionary matrix} is $\Pb_{\Tau_{N_T}}$. We wish to find a subset $\hat{\Tau}\subset\Tau_{N_T}$ with $\hat{L}=|\hat{\Tau|}$ such that 
1) the subspaces $\h_K\in\mathcal{S}(\F_{K,M}\Pb_{\{\tau_\ell\}_{\ell=1}^{L}})$ and $\h_S\in \mathcal{S}(\F_{K,M}\Pb_{\hat{\Tau}})$ are similar in the sense that $|\h_{E}|^2$ is small, and 2) the submatrix $\Pb_{\hat{\Tau}}\subset \Pb_{\Tau_{N_T}}$ is tall and thin so that its pseudoinverse exists and $|\tilde{\h}_S|^2\propto \frac{\hat{L}}{N}$ is small. 

Ideally, we would minimize the error directly solving
\begin{equation}
\label{eq:besterror}
\min_{\hat{\Tau}\subset\Tau_{N_T}}\min_{\bb}\Ex{\z}{|\h_E|^2+|\tilde{\h}_S|^2},
\end{equation}
where increasing $|\hat{\Tau}|$ increases $|\tilde{\h}_S|^2$ and decreases $|\h_E|^2$. However, $|\h_E|^2$ cannot be evaluated without knowing $\h_K$.

Instead we consider an approximate minimization for the error. To do so we first define
$$\h_{N/K}=\F_{N/K,M}\Pb_{\hat{\Tau}}\bb=\F_{N/K,M}\Pb_{\Tau_{N_T}}\bb_{N_T}=\Phib_{N_T}\bb_{N_T}$$
where $\bb$ is a non-sparse size-$\hat{L}$ vector and $\bb_{N_T}$ is the sparse size-$N_T$ vector with the coefficients of $\bb$ in the appropriate places and zeros elsewhere. If we define the matrix $\Phib_{N_T}=\F_{N/K,M}\Pb_{\Tau_{N_T}}$, then identifying the non-zero coefficients of the sparse vector $\bb_{N_T}$ is a classic CS problem. As an approximation to \eqref{eq:besterror} we use the $\ell$-0 minimization

\begin{equation}
\label{eq:bestsparse}
 \min |\hat{\bb}_{N_T}|_0 \textnormal{ s.t. } |\y_N-\Phib_{N_T}\hat{\bb}_{N_T}|^2\leq \xi.
\end{equation}
so that $|\hat{\Tau}|$ is minimized while $|\h_E|^2$ is constrained by the design parameter $\xi$.
Still, this $l$-0 problem has combinatorial complexity in the number of columns of $\Phib_{N_T}$. The CS literature can be roughly divided into two branches to address this: sufficient conditions on $\Phib$ such that some tractable problem is equivalent ($l$-1, LASSO, Dantzig-Selector), and heuristic approximations of \eqref{eq:bestsparse} \cite{Duarte2011}.

We choose the OMP algorithm as a heuristic approximation to our problem. OMP is a greedy algorithm that, on each iteration, adds one column to the matrix minimizing the LS projection of $\y_N$. Thus OMP has an interpretation as a heuristic for \eqref{eq:besterror} as well: each ``near sighted'' iteration follows the steepest decrease of $|\h_E|^2$ and increases $|\tilde{\h}_S|^2$, and OMP stops ($\xi$) when the next decrement of $|\h_E|^2$ is not worth its associated increase of $|\tilde{\h}_S|^2$. Algorithm \ref{alg:OMP} is our OMP with Binary-search Refinement where setting $\mu^*=0$ in line 8 converts it to the ``classic'' OMP algorithm.

 \begin{algorithm}[b]
\caption{Orthogonal Matching Pursuit with Binary-search Refinement (OMPBR)}
\label{alg:OMP}
\begin{algorithmic}[1]
  \STATE Define dictionary $\Tau_{N_T}=\{n\frac{D_s}{N_T}\}\forall n\in\{0\dots N_T-1\}$
  \STATE Generate $N/K$-FFTs  $\vv_n=\F_{N/K,M}\pp(\hat{\tau}_n)$
  \STATE Initialize estimate of sparse support set $\hat{\Tau}_0=\emptyset$
  \STATE Initialize residual with data observation $\rr_0=\y_{N/K}$
  \WHILE{$||\rr_i||^2>\xi$ and $i<$ max num. iterations}    
    \STATE $\overline{\tau}_i=\{\arg\max \vv_n^H\rr_{i-1}\forall \Tau_{N_T}\setminus\mathcal{T}_{i-1}\}$
    \STATE $\mu^*={\displaystyle \arg\max_{\mu \in[\frac{-1}{2},\frac{1}{2}]}} \pp(\overline{\tau}_i+\mu \frac{D_s}{N_T})^H\F_{N/K,M}^H\rr_{i-1}$
    \STATE $\hat{\tau}_i=\overline{\tau}_i+\mu^* \frac{D_s}{N_T}$
    \STATE Update estimation of support $\hat{\Tau}_{i}=\hat{\Tau}_{i-1}\cup \{\hat{\tau_i}\}$
    \STATE Update support matrix $\hat{\Phib}_i=\F_{N/K,M}\Pb_{\hat{\Tau}_{i}}$
    \STATE Update LS channel estimator  $\hat{\h}_i=\hat{\Phib}_i\hat{\Phib}_i^\dag\y_{N/K}$
    \STATE Update residual for next step $\rr_{i}=\y_{N/K}-\hat{\h}_i$, 
  \ENDWHILE
\end{algorithmic}
\end{algorithm}

By definition when $N_T=M$ the delay dictionary contains the sampling instants $\Tau_{M}=\{0,T,2T\dots (M-1)T\}$ and is a \textit{complete} dictionary without superresolution. On the other hand when $N_T>M$, $\Tau_{N_T}$ is an \textit{overcomplete} dictionary with superresolution. Unfortunately in the general case the matrix $\Phib_{N_T}$ would not meet sufficient conditions to guarantee that the representation of $\h_{N/K}$ as $\Phib_{N_T}\bb_{N_T}$ is unique \cite{Duarte2011}. Thus in CS OFDM channel estimation we cannot have simultaneously superresolution and invoke \cite[Theorems 6,7]{Duarte2011} to guarantee that the indices of the large coefficients of $\bb_{N_T}$ are recovered correctly. Nonetheless, in channel estimation particularly we do not care if $\hat{\Tau}$ contains false-positive errors as long as $|\tilde{\h}_K|^2$ is minimized.



By definition $\tau_\ell$ can take any real non-negative value, and therefore any finite dictionary ($N_T<\infty$) would incur some delay discretization error. It is common to disregard this error in mmWave CS channel estimator designs \cite{7953407,rodriguez2017frequency,Venugopal2017}, but in this paper we introduce the OMPBR modification to enable an effectively infinite dictionary $N_T=\infty$.

In summary we distinguish three cases
\begin{itemize}
 \item $N_T=M$ and $\mu^*=0$ is OMP with a finite dictionary without superresolution as in \cite{Taubock2008,Berger2010a,Qi2011a,7953407}. In this case, if the columns of $\Phib_M$ are orthogonal, for example if $p(t)$ is a Nyquist pulse, then greedy OMP solves the $l$-0 problem \eqref{eq:bestsparse} \textbf{exactly}.
 \item Finite $N_T> M$ and $\mu^*=0$ is OMP with a finite \textit{overcomplete} dictionary with superresolution as in \cite{Berger2010a,Venugopal2017}. This is an approximation of the $l$-0 problem  \eqref{eq:bestsparse}, so the error is lower bounded by the $l$-0 optimum for the same $N_T>M$. Even if we cannot invoke  \cite[Theorems 6,7]{Duarte2011}, a greedy algorithm with an extended dictionary always performs better, and therefore the error must be lower or equal than in OMP with $N_T=M$, which is also the $l$-0 optimum for $N_T=M$.
 \item Finite $N_T\geq M$ with $\mu^*$ as in Alg. \ref{alg:OMP} leads to our OMP with Binary-search Refinement which achieves an \textit{effective} infinite dictionary size. For this, lines 6-8 of Alg. \ref{alg:OMP} are equivalent to the subproblem
  $\max_{\tau} \pp(\tau)^H\F_{N/K,M}^H\rr_{i-1}$
 which is non-concave in the interval $[0,D_s]$. Conventional OMP addresses this by constraining $\tau$ to a finite dictionary. Instead, since typical pulses $p(t)$ such as the Raised Cosine are symmetric at $t=0$ and concave in the interval $-T/2,T/2$, we make the assumption that this problem is locally concave and symmetric around a local maximum in small regions we call \textit{delay bins}. First a finite dictionary is used to identify the best bin as in conventional OMP, centered at $\overline{\tau}_i$ with width $\pm\frac{1}{2}\frac{D_s}{N_T}$. Then the delay is refined as $\hat{\tau}_i=\overline{\tau}_i+\mu^* \frac{D_s}{N_T}$ where $\mu^*$ comes from an assumed locally concave and symmetric maximization. We use a Binary-search that successively divides the interval in halves like a bisection algorithm, rather than a gradient, to guarantee that OMPBR is robust in the sense that the result is contained in the bin and never worse than the decision that conventional OMP would have made.
\end{itemize}

\subsection{Asymptotic Error Analysis of OMP}
\label{sec:error}
We assume OMP stops after $\hat{L}$ iterations and returns $\Pb_{\hat{\Tau}}\in \mathbb{C}^{M\times\hat{L}}$. Both $\Pb_{\hat{\Tau}_{\hat{L}}}$ and $\hat{L}$ are random variables that depend on $\z_N$. The error variance can be written as
\begin{equation*}
\begin{split}
\nu_{OMP}^2&=\frac{\Ex{\z}{|\hat{\h}_K^{OMP}-\h_K|^2}}{K}\stackrel{a}{=}\frac{\Ex{\z}{|\tilde{\h}_S|^2+|\h_E|^2}}{K}\\
		&\stackrel{b}{=}\frac{\Ex{\z}{|\Pb_{\hat{\Tau}_{\hat{L}}}(\hat{\bb}-\bb)|^2+|(\I_M-\Pb_{\hat{\Tau}_{\hat{L}}}\Pb_{\hat{\Tau}_{\hat{L}}}^\dag)\h_M|^2}}{K}\\
\end{split}
\end{equation*}
where a) follows from the orthogonality of the vectors $\tilde{\h}_S$ and $\h_E$ and b) follows from $\F_{K,M}^H\F_{K,M}=\I_M$.

We define the random variable $\breve{\z}=\sqrt{\frac{K}{N}}\F_{N/K,M}^H\z_N$, where if $\z_N$ is AWGN, then $\breve{\z}$ is also AWGN with covariance $\sigma^2\I_M$. Using \eqref{eq:bML} the first error term is
$$\frac{\Ex{\z}{|\tilde{\h}_S|^2}}{K}=\frac{\Ex{\z}{\tr\{\Pb_{\hat{\Tau}}^H\breve{\z}\breve{\z}^H\Pb_{\hat{\Tau}}(\Pb_{\hat{\Tau}}^H\Pb_{\hat{\Tau}})^{-1}\}}}{N},$$
where $\Pb_{\hat{\Tau}}$ depends on $\breve{\z}$. We note that in the limit as $\sigma^2$ approaches zero the dependency vanishes, and thus
$$\frac{\Ex{\z}{|\tilde{\h}_S|^2}}{K}\stackrel{\sigma^2\ll1}{\xrightarrow{\hspace{.7cm}}}\frac{\sigma^2\tr\{\Pb_{\hat{\Tau}}^H\Pb_{\hat{\Tau}}(\Pb_{\hat{\Tau}}^H\Pb_{\hat{\Tau}})^{-1}\}}{N}=\frac{\Ex{\z}{\hat{L}}\sigma^2}{N}.$$

The second term is a decreasing function of $\hat{L}$ denoted
$$\rho(\hat{L})\triangleq \frac{|\h_E|^2}{K}=\frac{|(\I_M-\Pb_{\hat{\Tau}_{\hat{L}}}\Pb_{\hat{\Tau}_{\hat{L}}}^\dag)\h_M|^2}{K}.$$

OMP stops at the first iteration $i$ that satisfies $|\rr_{i}|^2\leq \xi$ where $\rr_{i}=(\I_N-\hat{\Phib}_i\hat{\Phib}_i^\dag)(\F_{N/K}\h_M+\z_N)$ as defined by lines 5 and 12 of Alg \ref{alg:OMP}. Thus we assume that the condition is approximately an equality for $i=\hat{L}$. With careful rearrangement of the factors of $\rr_{i}$ we can write 
{\small \begin{equation*}
  \Ex{\z}{\rho(\hat{L})}=\frac{\Ex{}{|\rr_{\hat{L}}|^2-|\z_N|^2+|\tilde{\h}_S|^2}}{N}\simeq\frac{\xi-N\sigma^2+\Ex{}{|\tilde{\h}_S|^2}}{N}
\end{equation*}}
%
Choosing $\xi=N\sigma^2$ gives $\Ex{\z}{\rho(\hat{L})}=\Ex{}{|\tilde{\h}_S|^2}/N$ and
{\small
\begin{equation}
 \label{eq:errOMP}
 \nu_{OMP}^2=\frac{2\Ex{}{|\tilde{\h}_S|^2}}{N}=\frac{2\Ex{}{\rho(\hat{L})}}{N}\stackrel{\sigma^2\ll1}{\xrightarrow{\hspace{.7cm}}}\frac{2\Ex{\z}{\hat{L}}}{N}\sigma^2.
\end{equation}}\\
For $M=N_T$ when $p(t)$ is a Nyquist pulse, it is easy to show that the high-SNR limit is a tight lower bound.
$$\nu_{OMP}^2\stackrel{M=N_T}{\geq}\frac{2\Ex{\z}{\hat{L}}}{N}\sigma^2.$$
In fact we have conjectured that this is a lower bound in the general case because the numerical results suggest so.

Additionally the analysis has shown that when OMP stops, the number of recovered MPCs, denoted by $\hat{L}$, is established at the meeting point between two terms, one that decays as $\rho(\hat{L})$ and one that grows linearly with $\hat{L}$
$$\Ex{\z}{\rho(\hat{L})}\simeq\frac{\Ex{}{|\tilde{\h}_S|^2}}{N}\stackrel{\sigma^2\ll1}{\xrightarrow{\hspace{.7cm}}}\frac{\Ex{\z}{\hat{L}}}{N}\sigma^2.$$
Here $\rho(d)$ is a random non-increasing function of $d$ with a decay that depends on the inequality between the coefficients of the vector $\h_M$. If we could characterize $\rho(d)$ in relation to the random distribution of $\h_M$, we would be able to use the stop-condition result in reverse to deduce $\Ex{\z}{\hat{L}}$ from $\h_M$ and evaluate $\nu_{OMP}^2$. Since this proves evasive, in the next section we propose a ``compressibility score'' of $\h_M$ to gauge the decay-speed of $\rho(d)$. We show this metric is loosely related to the fourth moment of the distribution of $\{\alpha_\ell\}_{\ell=1}^{L}$. Therefore, the more heavy-tailed the MPC amplitude distribution, the faster $\rho(d)$ decays and the lower the OMP estimation error lower bound $\frac{2\Ex{}{\hat{L}}}{N}\sigma^2$.

We verified in simulations that $\hat{L}$ grows with SNR (Fig. \ref{fig:ntaps128}) and that $\nu_{OMP}^2$ converges to $\frac{2\Ex{\z}{\hat{L}}}{N}\sigma^2$ (Fig. \ref{fig:mse128}).

\section{Compressibility Analysis}
\label{sec:OMPstop}

From \cite{Gribonval2012} we know that compressibility is related to the fourth moment of the distribution. Inspired by this, we propose scoring the compressibility of channel vectors with arbitrary distributions using the Fairness Index (FI), which is a traditional metric of inequality in the scheduling literature \cite{Jain1984}. Since the FI is traditionally defined for non-negative resource allocations \cite{Jain1984}, we define the power FI for a complex channel vector $\h_M$ as follows:
\begin{equation}
 \label{eq:FIdef}
 \textnormal{FI}(\h_M)=\frac{(\sum_{n=0}^{M-1}|h_M[n]|^2)^2}{M(\sum_{n=0}^{M-1}|h_M[n]|^4)}.
\end{equation}
This definition of the FI has the following properties:
\begin{itemize}
 \item If exactly $L\in\{1\dots M\}$ elements of $\h_M$ are non-zero and have equal magnitude, the FI is $L/M$
 \item If $\vv$ is a size-$M$ i.i.d. vector with coefficients following a zero-mean distribution $f(v[n])$, then 
 $$\lim_{M\to\infty}\textnormal{FI}(\vv)=\frac{\Ex{\vv}{|v[n]|^2}^2}{\Ex{\vv}{|v[n]|^4}}=\frac{1}{\kappa(v[n])}$$
 where $\kappa$ is the \textit{kurtosis} of the distribution $f(v[n])$. 
 \item Generally the FI of $\h_M$ is strongly related to the FI of the set $\{\alpha_{\ell}\}_{\ell=1}^{L}$. In particular, if the delays $\{\tau_\ell\}_{\ell=1}^{L}$ are exact multiples of $T$ and $p(t)$ is a Nyquist pulse, then $\textnormal{FI}(\h_M)=\frac{L}{M}\textnormal{FI}(\ab)$. 
 \item The FI is invariant to scale, so 
 we can disregard the normalization  of the set $\{\alpha_{\ell}\}_{\ell=1}^{L}$ specified in \cite{Mathew2016,Specification2017}. $\textnormal{FI}(\h_M)$ is strongly related with the fourth moment of the lognormal distribution used to generate the unnormalized amplitudes $\{\overline{\alpha}_{\ell}\}_{\ell=1}^{L}$.
\end{itemize}



We now focus on the case where $N_T=M$ and $p(t)$ is a Nyquist pulse and show the connection between $\textnormal{FI}(\h_M)$ and $\rho(d)$. First, we define the sequence of magnitudes of the coefficients of $\h_M$ \textit{sorted in decreasing order} as
$$m_i=\begin{cases}
       \max (\{|h_M[n]|^2\}_{n=1}^{M})& i=1\\
       \max (\{|h_M[n]|^2\}_{n=1}^{M}\setminus \bigcup_{j=1}^{i-1}\{m_j\})& i>1\\
      \end{cases}
$$
where $m_1$ is the power of the largest coefficient, $m_2$ the second, and so on. Since $\sum_{\ell=1}^{L}\alpha_\ell^2=1$ and $|\pp(\tau)|^2=1$, we get $|\h_M|^2=1$ and $1-\sum_{i=1}^{d-1}m_i=\sum_{i=d}^{M}m_i$.

Second, assuming that after $\hat{L}$ iterations the set $\{m_i\}_{i=1}^{\hat{L}}$ is recovered perfectly, we define a recursive function that depends only on $\h_M$ and not on the noise as
\begin{equation}
\label{eq:defrhoid}
\overline{\rho}(d)\triangleq \frac{1}{K}\left(1-\sum_{i=1}^{d}m_i\right)=\overline{\rho}(d-1)\left(1-\frac{m_{d}}{\sum_{i=d}^{M}m_i}\right)
\end{equation}
where it is possible to verify by induction that $\rho(d)\geq\overline{\rho}(d)$, and in the high-SNR limit this becomes an equality.


Finally, we relate $\overline{\rho}(d)$ to the FI.
We first define a set containing the residual channel coefficient powers after the $d$ strongest are perfectly recovered, denoted by
$$\mathcal{R}_d\triangleq \{m_j\}_{j=d+1}^{M}=\{|h_M[n]|^2\}_{n=1}^{M}\setminus \{m_j\}_{j=1}^{d}.$$
The FI of this set evaluates to
$$\textnormal{FI}(\mathcal{R}_d)=\frac{(\sum_{i=d+1}^{M}m_i)^2}{(M-d)\sum_{i=d+1}^{M}m_i^2}.$$

By definition the FI satisfies the following inequalities
\begin{equation}
\label{eq:FIbounds}
\frac{1}{(M-d)\sqrt{\textnormal{FI}(\mathcal{R}_d)}}\leq\frac{m_d}{\sum_{i=d+1}^{M}m_i}\leq\frac{1}{\sqrt{(M-d)\textnormal{FI}(\mathcal{R}_d)}}
\end{equation}
Introducing the right hand side of \eqref{eq:FIbounds} into \eqref{eq:defrhoid} we get
\begin{equation}
\label{eq:rhoLB1}
\begin{split}
  \rho(d)\geq\overline{\rho}(d)
  &\geq\prod_{i=0}^{d-1}\left(1-\frac{1}{\sqrt{(M-i)\textnormal{FI}(\mathcal{R}_i)}}\right).
  \end{split}
\end{equation}

So we see that, if the $\textnormal{FI}(\mathcal{R}_i)$'s are high, $\overline{\rho}(d)$ cannot decay fast. The left hand side of \eqref{eq:FIbounds} can be used to write a converse (if $\textnormal{FI}(\mathcal{R}_i)$'s are low $\overline{\rho}(d)$ must decay fast) as
\begin{equation}
\label{eq:rhoUN}\overline{\rho}(d)\leq\prod_{i=0}^{d-1}\left(1-\frac{1}{(M-i)\sqrt{\textnormal{FI}(\mathcal{R}_i)}}\right).\end{equation}
We plot the empirical values of \eqref{eq:rhoLB1} in Fig. \ref{fig:boundsrho} (averaged over the random mmWave channel distribution $\Ex{\h_M}{.}$), where we see that it is a lower bound of $\overline{\rho}(d)$.

At this point we have bounds depending on the values of $\textnormal{FI}(\mathcal{R}_i)$ for $i\in[0,d-1]$, while we want a bound depending on $\textnormal{FI}(\h_M)=\textnormal{FI}(\mathcal{R}_0)$. To address this we introduce the following assumption: When $d\ll M-d$ we assume that with a high probability, removing the largest element from $\mathcal{R}_d$ increases the FI by a factor of at least $\frac{M-d+1}{M-d}$, i.e.
\begin{equation}
\label{eq:assumptionFI}
\textnormal{FI}(\mathcal{R}_d)\stackrel{M-d\gg d}{\gtrsim}\frac{M-d+1}{M-d}\textnormal{FI}(\mathcal{R}_{d-1})\;w.h.p.
\end{equation}
This approximation was tested numerically and seems to hold with a high probability for mmWave channels. Using this approximation we can recursively replace $\textnormal{FI}(\mathcal{R}_d)$ with $\textnormal{FI}(\mathcal{R}_{d-1})$ until $FI(\h_M)=\textnormal{FI}(\mathcal{R}_0)$ to write an approximation of \eqref{eq:rhoLB1} as the following geometric progression
{\small \begin{equation}
\label{eq:rhoLB2}
\overline{\rho}(d)\gtrsim\left(1-\frac{1}{\sqrt{M \textnormal{FI}(\h_M)}}\right)^d.
\end{equation}}
This approximation shows how $\textnormal{FI}(\h_M)$ influences the ``compressibility'' of the vector $\h_M$: the lower $\textnormal{FI}(\h_M)$ is, the faster $\overline{\rho}(d)$ decays, and the sooner OMP meets its stop condition, retrieving fewer channel MPCs $\Ex{\z}{\hat{L}}$. Since the error is $\sim\frac{2\Ex{\z}{\hat{L}}}{N}\sigma^2$, this concentration of channel power in a few dominant MPCs is desirable.
We represented the empirical values of \eqref{eq:rhoLB2} in Fig. \ref{fig:boundsrho}, where we see that it is a very close approximation of \eqref{eq:rhoLB1} and a lower bound for empirical values of $\rho(d)$ obtained in simulation.

\section{Numerical Results}
\label{sec:num}

We simulate a mmWave OFDM system with $T=2.5$ns and $p(t)=\textnormal{sinc}(t/T)$. This makes the bandwidth $B=400$MHz. The DFT size is $K=512$ and the CP length $M=128$, for a maximum delay spread $D_s=TM=320$ns. We assume the channel is time-invariant over several OFDM frames, where the first frame contains $N=128$ pilots. Results are computed over $10^3$ independent channel MPC realizations generated by the NYUWireless model \cite{Mathew2016} for a $100$m non-line-of-sight link with $f_c=28$GHz.

\begin{figure}
\centering
 \includegraphics[width=.9\columnwidth]{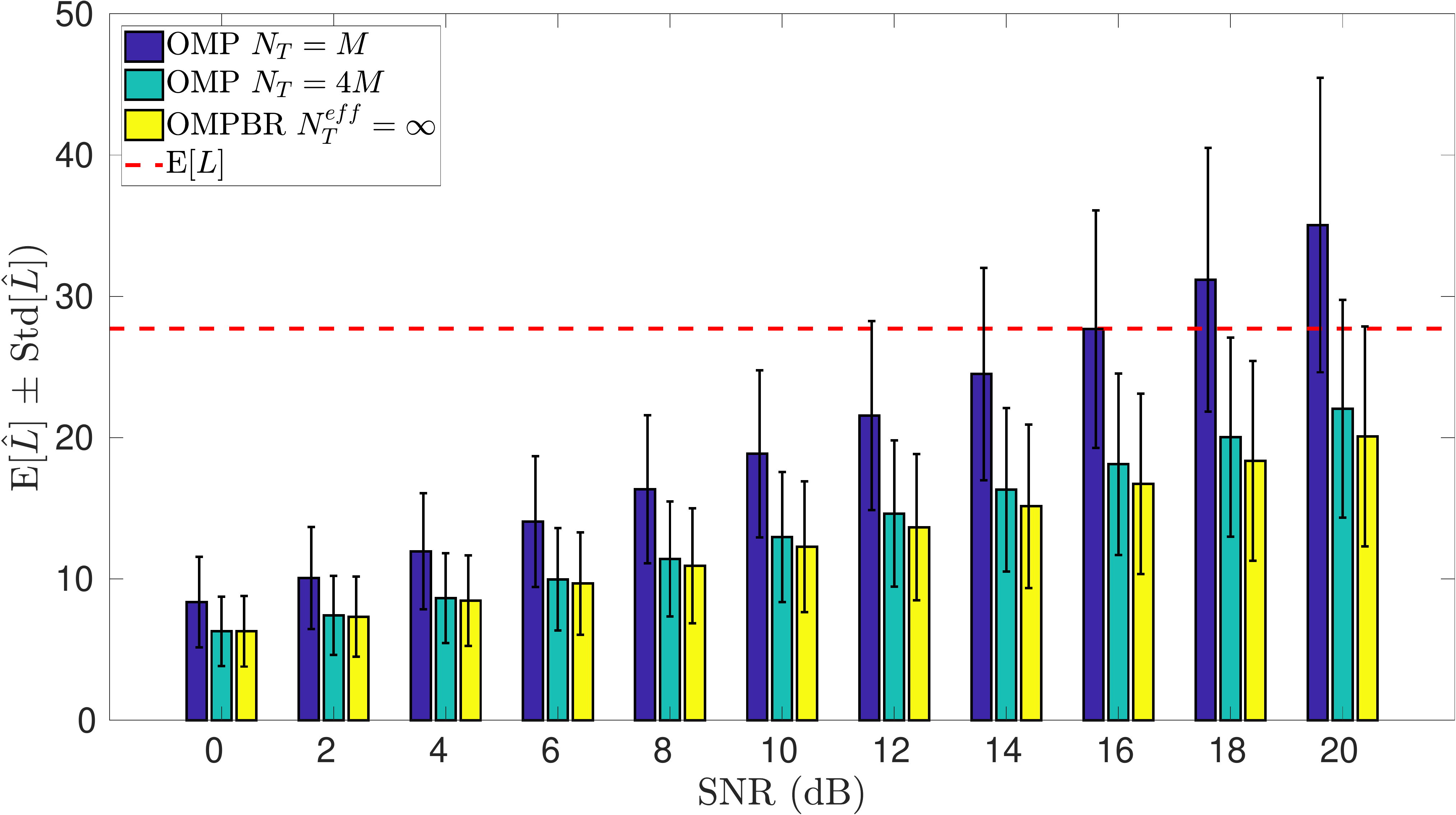}
 \caption{Mean value of $\hat{L}$ and standard deviation for OMP vs SNR}
 \label{fig:ntaps128}
\end{figure}

First we look at $\hat{L}$. In Fig \ref{fig:ntaps128} we represent $\Ex{\z,\h_M}{\hat{L}}$ vs SNR for three estimators: no-superresolution OMP ($N_T=M$), conventional superresolution OMP ($N_T=4M$), and our OMP Binary-search Refinement proposal ($N_T$ is effectively infinite). The channel model \cite{Mathew2016} generates $L$ as a random number that does not depend on SNR, and the average in our simulations was $\Ex{}{L}=28$, represented by a red line in the figure. The bar plots show the number of MPCs retrieved by OMP, which grow as SNR increases consistent with our analysis.  At low SNR no-superresolution OMP estimates about 8 MPCs, whereas both algorithms with superresolution start at 5. At high SNR orthogonal OMP overshoots to 35 due to its insufficient delay resolution, whereas the dictionaries with superresolution estimate about 20 dominant MPCs. As we have argued in Section \ref{sec:OMPstop}, for practical SNR values a number of MPCs of the mmWave channel are too weak to be recovered by CS techniques and $\hat{L}<L$.

%

\begin{figure}
 \centering
 \includegraphics[width=.9\columnwidth]{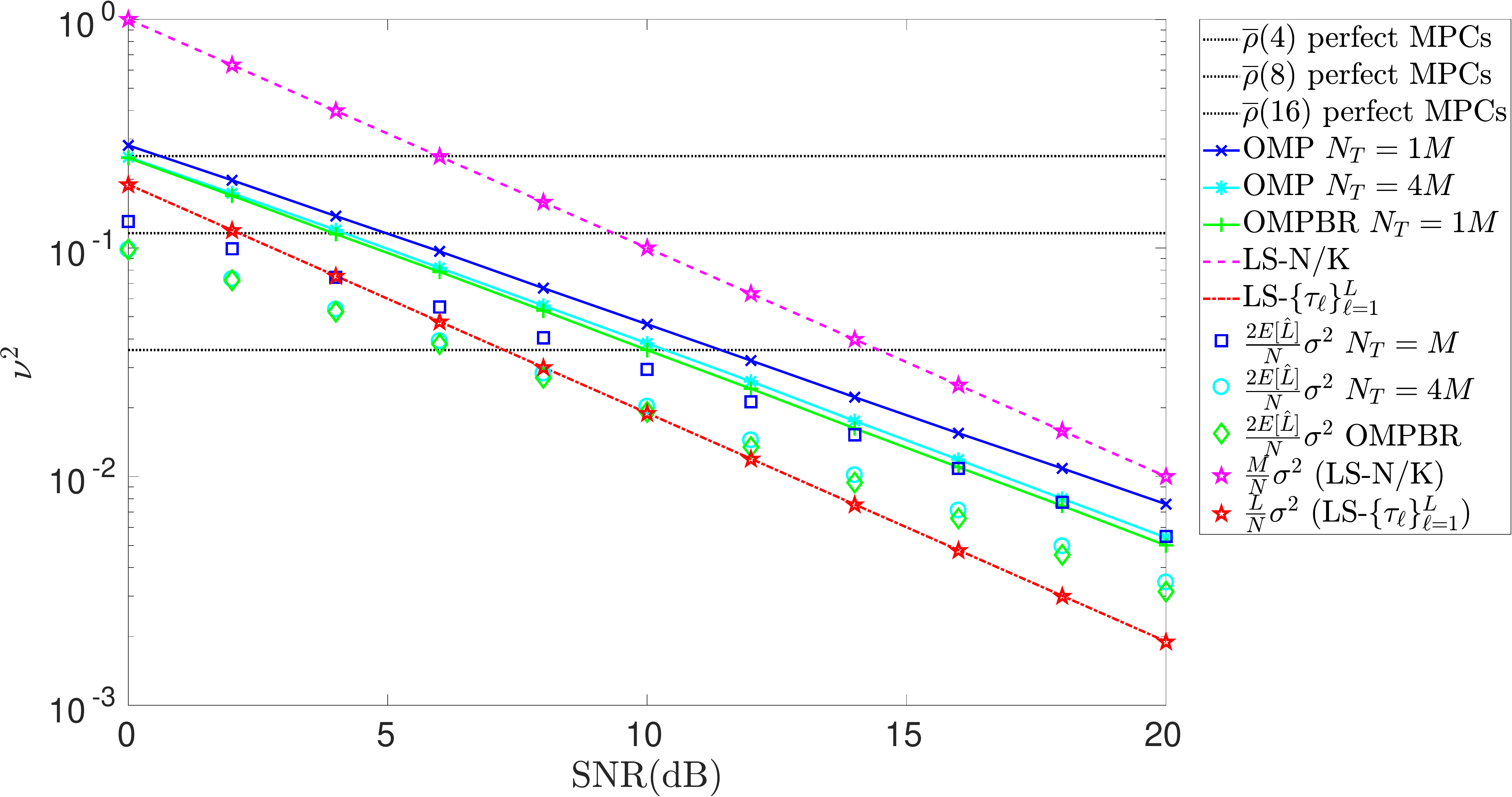}
 \caption{Channel estimation error variance ($\nu^2$) vs SNR.}
 \label{fig:mse128}
\end{figure}

%
Next we verify the results for the estimation error in Fig. \ref{fig:mse128}. The purple line corresponds to the error of the non-sparse LS-ML estimator benchmark defined in Sec. \ref{sec:LSMLconventional}. Its numerical result coincides with our theoretical prediction \eqref{eq:errMLMK}, represented by star-shaped purple bullets in the figure. We represent with a red line the genie-aided LS-ML sparse estimator benchmark defined in Sec. \ref{sec:LSMLMPC}, where we see again the error coincides with our theoretical prediction  \eqref{eq:errMLaK} (red star bullets). Since $\Ex{\h_M}{L}=28$ and $N=M=128$, the sparse benchmark has a gain of approximately $6$dB.
Next, we have the OMP algorithm with $N_T=M$,  $N_T=4M$, and OMPBR with an effectively infinite dictionary. The error variance analysis in Sec. \ref{sec:CS} predicts the error converges to $\frac{\Ex{\z,\h_M}{2\hat{L}}}{N}\sigma^2$ in the high-SNR regime. We represent the empirical error with solid green, cyan and blue lines, and we represent with squared, circular and diamond bullets of the same colors the theoretical lower bound of the error evaluated taking the empirical values of $\Ex{\z,\h_M}{\hat{L}}$ previously displayed in Fig \ref{fig:ntaps128}. We confirm that the theoretical results approximate the empirical error from below and become tighter as the SNR increases. The CS estimators outperform the non-sparse estimator benchmark but fail to match the genie-aided sparse estimator benchmark. Particularly at low SNRs OMP displays a lower $\Ex{\z,\h_M}{\hat{L}}$ in Fig. \ref{fig:ntaps128} and achieves a greater advantage over the non-sparse benchmark.


\begin{figure}
 \centering 
 \includegraphics[width=.9\columnwidth]{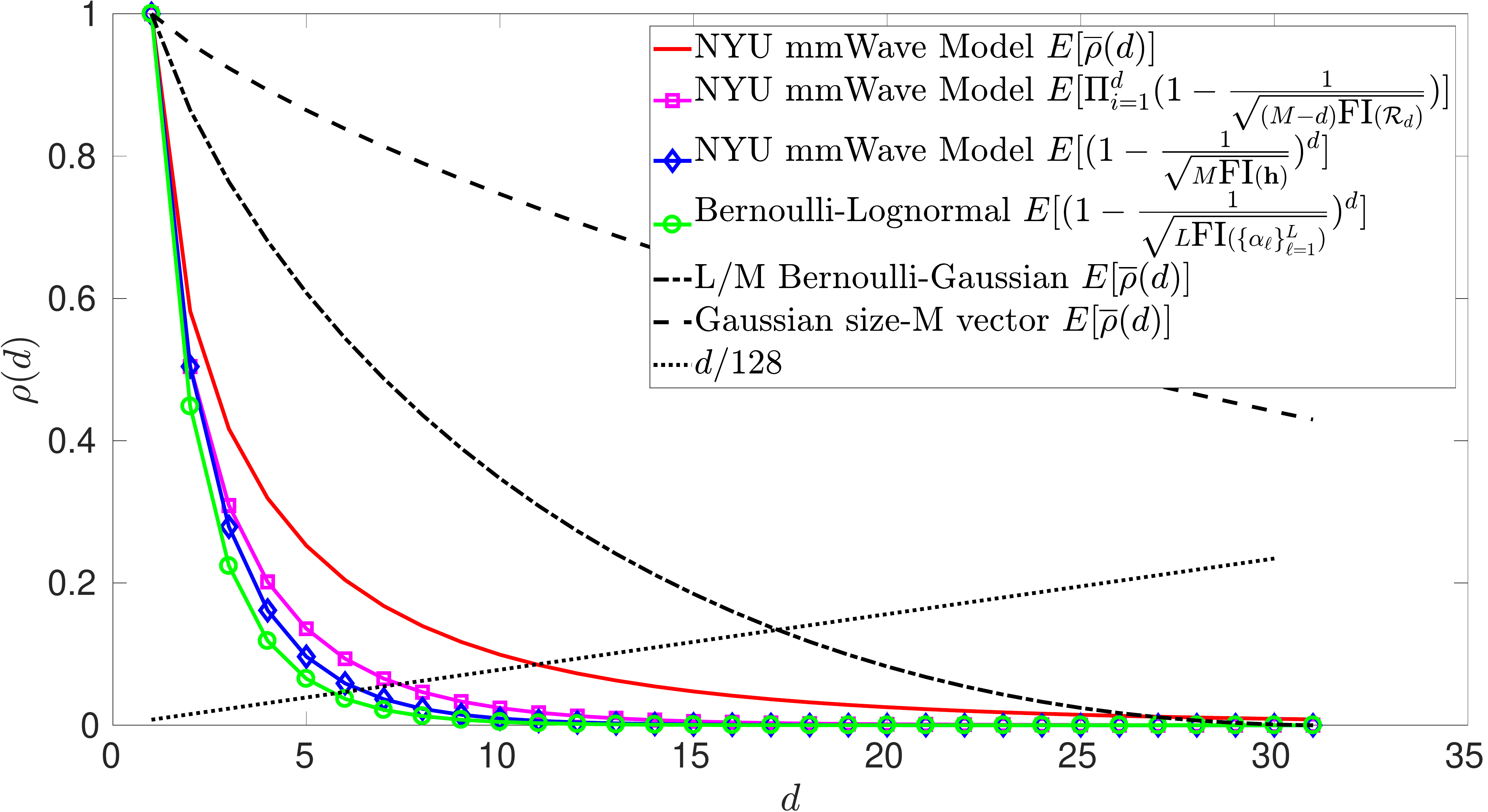}
 \caption{Comparison of $\Ex{\h_M}{\overline{\rho}(d)}$ for a mmWave channel, sparse and non-sparse Gaussian channels, and the bounds obtained in Sec. \ref{sec:OMPstop}.}
 \label{fig:boundsrho}
\end{figure}

Finally we display the relation between the OMP algorithm and the ``compressibility'' of the vector $\h_M$ represented by the function $\rho(d)$ and the Fairness Index of $\h_M$. We depict the empirical average of the function $\Ex{\h_M}{\overline{\rho}(d)}$ versus $d$ for mmWave channels using a solid red line in Fig. \ref{fig:boundsrho}. We observe that this function decays exponentially with $d$. We also depict the lower bounds \eqref{eq:rhoLB1} and \eqref{eq:rhoLB2} using purple and blue lines, respectively. The green line represents a Bernoulli-Lognormal simplified-mmWave channel model, where $\textnormal{FI}(\h_M)$ is exactly $\frac{L}{M}\textnormal{FI}(\{\alpha_\ell\}_{\ell=1}^{L})$. Finally the dash-dotted and dashed black lines represent $\overline{\rho}(d)$ for Bernoulli-Gaussian and non-sparse Gaussian vectors, showing that for channels generated by these non-heavy-tailed distributions, the function $\overline{\rho}(d)$ decays much more slowly and OMP would have to run for many more iterations or endure a higher error variance in the channel estimation. A reference black dotted line in Fig. \ref{fig:boundsrho} helps to illustrate how the stop condition of OMP works. The values of $d$ where this monotonously-increasing line meets each of the decaying functions provide an approximate indicator of the number of iterations that OMP would run.

\section{Conclusions}
\label{sec:conclusions}

We have proposed CS estimators and analyzed the ``compressibility'' (i.e. CS performance under a random vector generator) for wireless channels with a small number of MPCs. We obtained an analytical approximation for the error of the OMP algorithm as a function of the number of iterations, $\frac{\Ex{}{2\hat{L}}}{N}\sigma^2$. We show $\Ex{}{\hat{L}}$ depends on the decay speed of a certain decreasing function $\rho(d)$ that captures the inequality in magnitude of the coefficients of the vector. Inspired by an existing result that connects the fourth moment and the compressibility of random i.i.d. vectors, we have used the Fairness Index to score the compressibility of arbitrary channel vectors. We have shown analytically that the lower the FI of the channel, the faster $\rho(d)$ decays, and hence the lower the error variance of the OMP channel estimator. Measurements of mmWave channels are consistent with a heavy tailed lognormal distribution. Thus, in addition to being physically sparse in the sense that the number of MPCs is lower than the number of channel taps, mmWave channels are statistically sparse in the sense that the set of MPCs has a very low FI. Therefore our results show that measurement-based mmWave channel models are even better suited for CS estimation than the Bernoulli-Gaussian distributions assumed in prior works.

\end{document}